# Recognizing Salt Wave Events in Coastal Systems

**Xun Cai[1], Qubin Qin[2], Matthew Kirwan[3], Holly Michael[4], Jian Shen[3], Katharine J. Mach[5], Peter Raymond[1]**

[1]School of the Environment, Yale University, New Haven, CT, 06511, USA.

[2]Coastal Studies Institute, East Carolina University, Wanchese, NC, 27981, USA.

[3]Virginia Institute of Marine Science, William & Mary, Gloucester Point, VA, 23062, USA

[4]Department of Earth Sciences, University of Delaware, Newark, DE, 19716, USA

[5]Rosenstiel School of Marine, Atmospheric, & Earth Science, University of Miami, Miami, FL, 33149, USA

Corresponding author: Xun Cai (xun.cai@yale.edu); ORCID: 0000-0002-4251-2384

**Key Points:**

- We introduce the concept of "coastal salt waves" – periods of elevated salinity anomalies – to characterize extreme salinity events.
- Coastal salt waves exhibit significant spatial and temporal variability based on duration, intensity, frequency, cumulation, and timing.
- Drivers and and resulting ecological disturbance of coastal salt waves may differ from gradual saltwater intrusion.

**Abstract**

Saltwater intrusion is a critical challenge to coastal ecosystems, impacting freshwater resources, biogeochemical cycles, and habitat stability. While relevant studies often focus on the long-term trends of salinity, its episodic variability and resulting ecological disturbance remains underexplored. Here, we introduce the concept of "coastal salt waves" – periods of elevated salinity anomalies, akin to heat waves – to better characterize extreme salinity events and emphasize their significance. Using cases studies, we show that coastal salt waves exhibit significant spatial and temporal variability based on their duration, intensity, frequency, cumulation, and timing, with drivers and impacts that may differ from gradual saltwater intrusion. In years with similar average salinity, salt waves may still vary greatly in characteristics like intensity, resulting in varying environmental impacts. Furthermore, systems without rising average salinity may still face more frequent or intense salt waves. This framework supports monitoring and management strategies to mitigate coastal salinization risks.

**Plain Language Summary**

Saltwater intrusion is an emerging challenge in coastal ecosystems, threatening drinking water supplies, biodiversity, agriculture, and habitat stability. While long-term trends are well-studied, short-term extreme salinity events, lasting days to months, remain underexplored despite their potential to cause substantial ecological stress, such as triggering ecosystem shifts, vegetation loss, soil chemistry alterations, and habitat disruption. The concept of coastal salt waves, defined as prolonged high-salinity anomalies, presents a framework to quantify their durations, intensities, frequencies, cumulations, and timings. By integrating salt wave metrics with observational data and modeling, this framework improves the ability to assess and manage salinity-driven changes in coastal ecosystems amid climate variability and human influences.

# 1 Introduction

Coastal systems, including estuaries, tidal rivers, intertidal zones, and groundwater aquifers, serve as critical transition zones between oceanic and freshwater environments. Many coastal systems experience substantial spatial and temporal salinity variability (Werner et al., 2013; Michael et al., 2013), which plays a crucial role in shaping coastal water resources and ecosystem structure. Increased salinity in river surface waters poses risks to drinking water supplies and quality, while salinization in aquifers can hinder agricultural irrigation (Barlow and Reichard, 2010; Werner et al., 2013; Silva et al., 2015). Salinity is also a key regulator of ecosystem productivity, nutrient cycling, and organism performance (Velasco et al., 2018). For example, salinity patterns dictate the dominant algal communities and affect microbial processes such as sulfate-driven organic matter decay in shallow water systems (Yukoi et al., 2002; Berger et al., 2019). Overall, given the critical role that salinity plays in coastal systems, understanding salinity variations – especially in the context of climate change, demands greater attention.

A critical coastal environmental issue is the intrusion of seawater into freshwater surface zones, coastal aquifers, and other coastal systems (Barlow and Reichard, 2010; Werner et al., 2013), often referred to as saltwater intrusion (SWI). SWI drives ecological shifts, such as coastal forest die-offs that lead to shifts in biodiversity (Kirwan and Gedan, 2019; Tully et al., 2019). And SWI poses a critical threat to vegetation species lacking the physiological adaptations to tolerate such conditions (Yukoi et al., 2002; Munns and Tester, 2008; Muylaert et al., 2009; Anderson et al., 2022). In addition, increased salinity may trigger accelerated decomposition of organic matter and destabilization of soil structure, leading to peat collapse and subsequent land subsidence (Chambers et al., 2019; Kirwan et al., 2024). These changes further increase the vulnerability of coastal landscapes by reinforcing feedbacks, such as increased rates of shallow subsidence and elevated tree mortality, which together contribute to the degradation of coastal ecosystems. (Kirwan et al., 2024).

SWI may be a more severe threat to coastal ecosystems in the era of climate change as sea-level rise (SLR) – driven by ocean warming and the melting of glaciers and ice sheets – intensifies the threat. Using the metrics of annual or monthly means, salt front distance, and salt fluxes (Barlow and Reichard, 2010; Carvine et al., 1992; Geyer et al., 2020; Liu et al., 2024), numerous studies have suggested that enhanced SWI has emerged as a significant trend along with SLR (Kirwan et al., 2024). However, SWI is influenced by multiple processes including shifts in water table elevations, rapid processes such as storms, tides, or winds regulate the intensity, duration, and frequency of SWI on shorter timescales (Tully et al., 2019), as documented by associated substantial fluctuations in salinity or specific conductivity at local monitoring stations (Fig. 1).

The ecological effects on ecosystems resulting from these disturbances by these salinity anomalies may also be as important as the long-term trends of SWI. For example, a high-intensity (e.g., 4 dS/m for crops) salinity event can significantly contribute to osmotic stress, disrupting cellular processes and resulting in the rapid die-off of freshwater plants or alterations in fish habitats (Barletta et al., 2005; Munns and Tester, 2008). Alternatively, a prolonged duration of mild intensity event can result in chronic stress, ultimately causing changes in productivity and ecological functions of tidal marsh systems (Janouse and Mayo, 2013; Neubauer, 2013). Thus, including analysis of these salinity anomalies, especially episodic and extreme events, along with the long-term trends of SWI is needed for a more complete quantification of SWI as well as its effects on coastal ecosystems. These anomalies, however, have not been adequately captured previously and there remains a lack of a framework for quantifying SWI based on the potential environmental disturbances. To address this gap, here we review the impacts and drivers of salinity anomalies and propose establishing specific criteria within the framework of "coastal salt waves" to emphasize the role of salinity anomalies across the coastal zone.

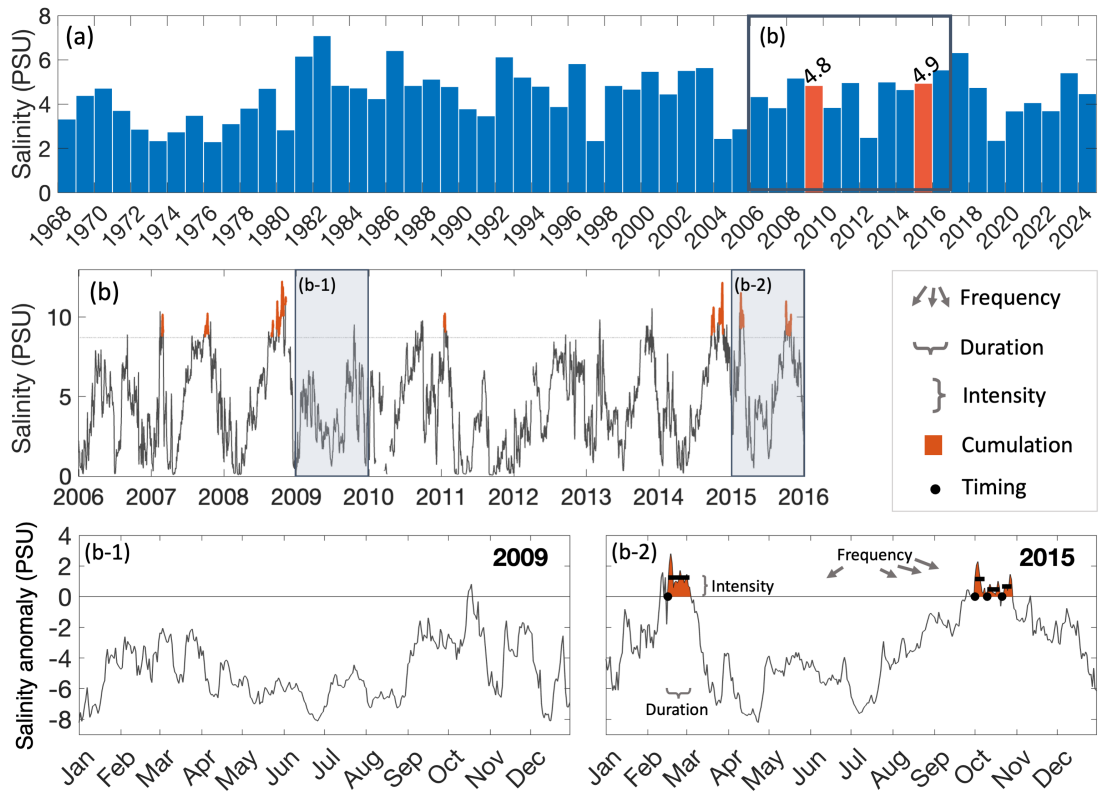

**Figure 1**. (a) Annual average surface water salinity measured at USGS station 01482800. (b) Daily salinity time series with detailed views in panels (b-1) and (b-2), showcasing two selected years with similar annual averages, marked by red bars in panel (a). Salt wave events are highlighted in panel (b-2), where red polygons delineate the events, with each polygon's area representing the cumulative magnitude of the respective event.

**2 Definition and characteristics of coastal salt waves**

Similar to the well-established concept of marine heat waves (Hobday et al., 2016), we define coastal salt waves as periods of consecutive days when salinity indices are abnormally high. Specifically, we identify salt waves as prolonged salinity events lasting more than a given number of days, where local salinity indices exceed a defined threshold. Clearly, different critical durations and thresholds can be selected depending on the research focus or locally well-established biological criteria. In this manuscript, for the sake of simplicity and generality, we adopt the marine heat wave definition by setting the critical duration of 5 days (Hobday et al., 2016). Note that rather than removing the seasonal cycle from the salinity data, based on all available datasets with a monitoring history of at least 15 years, as opposed to the 90th percentile of climatology adopted by heatwaves because salinity exhibits more variations than temperature, and its regional variations over latitudes are not as pronounced. This approach enables the definition of coastal salt waves in a way that is more relevant to ecosystem impacts while preserving absolute salinity thresholds.

We introduce a number of indices that may be important with respect to impacts on different system components to evaluate salt wave events over time: (1) duration, (2) intensity, (3) frequency, (4) cumulation, and (5) timing (Table 1). Drawing on the multiple ecosystem services outlined in the Millennium Ecosystem Assessment, we reviewed a few key processes (Table 2) that are associated with these salt wave events and are likely to alter ecosystem functions, as described by Talbot et al. (2018).

Duration refers to the consecutive days a salt wave event persists. The duration of salt waves, particularly mild ones, is closely associated with chronic stressors. For instance, a prolonged yet moderate increase in salinity can lead to long-term ecological effects, such as reduced vegetation productivity (Neubauer, 2013) and shifts in dominant plankton communities (Muylaert et al., 2009). These changes in vegetation and plankton communities, in turn, influence essential ecosystem functions like productivity and nutrient cycling (Table 2). While longer durations are often linked to greater intensity (Supplement Fig. 2), duration primarily emphasizes the impacts of salinity anomaly over time.

Intensity refers to the extent to which salinity anomalies exceed a defined threshold. Higher intensity events (i.e., a larger divergence from a threshold) are directly linked to osmotic stress, resulting in immediate impacts such as rapid vegetation die-offs and long-term consequences like the formation of ghost forests (Munns and Tester, 2008; DiCara and Gedan, 2023; Lorrain-Soligon et al., 2024). These vegetation die-offs not only contribute to landscape degradation but also diminish the ecosystem's aesthetic value and functional integrity (Table 2). Thus, intensity plays a critical role in shaping both the immediate biological responses and the long-term ecological resilience of the coastal regions.

Frequency represents the number of occurrences per year. While duration and intensity are crucial for assessing the impacts of single events, frequency and cumulation (see below) provide insight into impacts on local salt conditions on longer timescales. A high frequency of salt wave events intensifies the effects of elevated salinity exposure. For instance, frequent salt wave events trigger more frequent releases of soil contaminants and therefore pollution levels can escalate rapidly (Ding et al., 2023), posing greater risks to drinking water supplies and food security. Therefore, frequent salt wave events tend to compound environmental stress, amplifying long-term risks to both ecosystem health and human well-being.

Cumulation is the sum of salinity anomalies over a year. By summing up the occurrence of duration, intensity, and frequency, cumulation provides a holistic view of salinity stress on ecosystem. The compounding impact of multiple salt wave events may be more significant than that of individual occurrences, leading to ecosystem changes such as mortality of organisms (Blakeslee et al., 2013) and shifts in microbial communities (Zhang et al., 2021). These shifts in microbial processes not only affect greenhouse gas emissions but also play a role in regulating aquatic diseases (Yukoi et al., 2002; Berger et al., 2019). Additionally, on land, the cumulative salinity stress influences soil erodibility, while in estuarine environments, the position of the salt front affects the turbidity maximum, thereby impacting sediment transport and delivery (Chambers et al., 2019; Schoellhamer, 2000). The impacts on soil erosion and sediment delivery further influence ecosystem functions related to soil formation and landscape development. Ultimately, cumulation captures the compounded and far-reaching consequences of repeated salt wave events, such as systemic ecosystem transformations across both terrestrial and aquatic environments.

Timing refers to the the seasonality of salt waves events, with phenological relevance for coastal organisms. The timing of salt wave events can significantly interact with the life history of many coastal organisms, such as phenology of phytoplankton production (Gasiunaite et al., 2005), survival of different life stages of organisms (Blakeslee et al., 2013), and fish larvae developments (Welch et al., 2019). Therefore the timing of salt waves further impacts primary production, water quality, fishery resources, and coastal ecosystems' food supplies (Table 2). Thus, the timing of salt wave events not only influences immediate biological responses but can also trigger cascading effects that shape the long-term structure and productivity of coastal ecosystems. While some of effects caused by the different characteristics of salt waves may be reversible, others can result in prolonged disturbances with lasting consequences.

**Table 1**. Metrics of salt waves and examples of environmental responses for each criterion.

| Metrics | Definition | Major drivers | Environmental response | Reference(s) |
| --- | --- | --- | --- | --- |

| | | | | | |
|---|---|---|---|---|---|
| Duration | Number of consecutive days a high-salinity event persists | o | Drought | Adapt or chronic stress: Long-term shift species in plankton communities; Mortality or decreasing production in mangroves and salt marshes | Muylaert et al. (2009); Janousek and Mayo (2013); Neubauer (2013) |
| | | o | Discharge | | |
| | | o | Precipitation | | |
| | | o | Evaporation | | |
| | | o | Storm surges | | |
| | | o | Hurricanes | | |
| Intensity | Average salinity anomaly exceeding the defined threshold | o | Groundwater overdraft | Osmotic stress: Die-off of freshwater vegetation species; Rapid shift of estuarine fish habitats | McDonald (1955); Barletta et al. (2005); Munns and Tester (2008) |
| | | o | River damming and diversion | | |
| | | o | Urbanization | | |
| | | o | Agricultural irrigation | | |
| Frequency | Number of high salinity events occurring annually | o | Coastal erosion | Enhanced impacts of salinity exposure; Corrosion in coastal infrastructure Release of soil contaminants | Kamer and Fong (2000); Zhang et al. (2022); Silva et al. (2015) |
| | | o | Sea-level rise | | |
| Cumulation | The product of its mean intensity and duration | | | Both chronic stress and osmotic stress | Blakeslee et al. (2013) |
| Timing | Starting day of each high salinity event within a given year | | | Change on life cycles of coastal organisms: survival of larval development; Phenology of phytoplankton production | Gasiunaite et al. (2005); Welch et al. (2019) |

**Table 2**. Ecosystem services and their associated indicators for assessing changes, along with their connection to salt wave events.

| Ecosystem service | Indicator | Process related to salt wave events |
|---|---|---|
| Primary production | Primary production | Change nutrients supply, productivity of primary producers, and dominant species |
| Soil formation | Erosion and deposition | Change soil erosion and location of turbidity maximum |
| Water quality | Nitrogen and phosphorus concentration | Change nutrient cycling from production and metabolism |
| Climate regulation | Greenhouse gas release | Bring sulfate to anaerobic microbial process on the decomposition of organic matter |
| Drinking water | Chloride concentration | Decline freshwater availability and increase water treatment cost |
| Food supply | Crops harvest and fish catch | Decrease farmland area and change fish habitat |
| Disease regulation | Microbial activities | Change microbial growth and distribution |
| Aesthetic value | Housing value discount | Degrade landscape and weaken structural integrity due to corrosion of building materials |

## 3 Case studies

### 3.1 Frequent salt waves can occur during moderate salinity years

We evaluated a series of case studies to demonstrate the variations and influencing factors using literature and available surface water salinity data along the U.S. mid-Atlantic and parts of the New England Coast. Salt wave metrics' temporal variability differs significantly based on salinity trends and across salt wave metrics (Figs. 1 and 2), as also supported by heatmaps in Supplement Fig. 1. Importantly, when annual salinity levels are similar across different years, salt wave events can vary considerably in their timing, intensity, and duration (Fig. 1). While saltier years generally have a higher likelihood of experiencing more frequent or stronger salt waves, intense or frequent salt waves can also occur during mildly salty years. For example, in 2007 in the lower Chesapeake Bay, annual salinity was around the 50th percentile of the historical record, yet the frequency of salt wave events was at its highest (Supplement Fig. 1). These findings underscore that salt wave dynamics are not solely governed by average salinity levels, highlighting the need to consider event-scale variability on coastal systems.

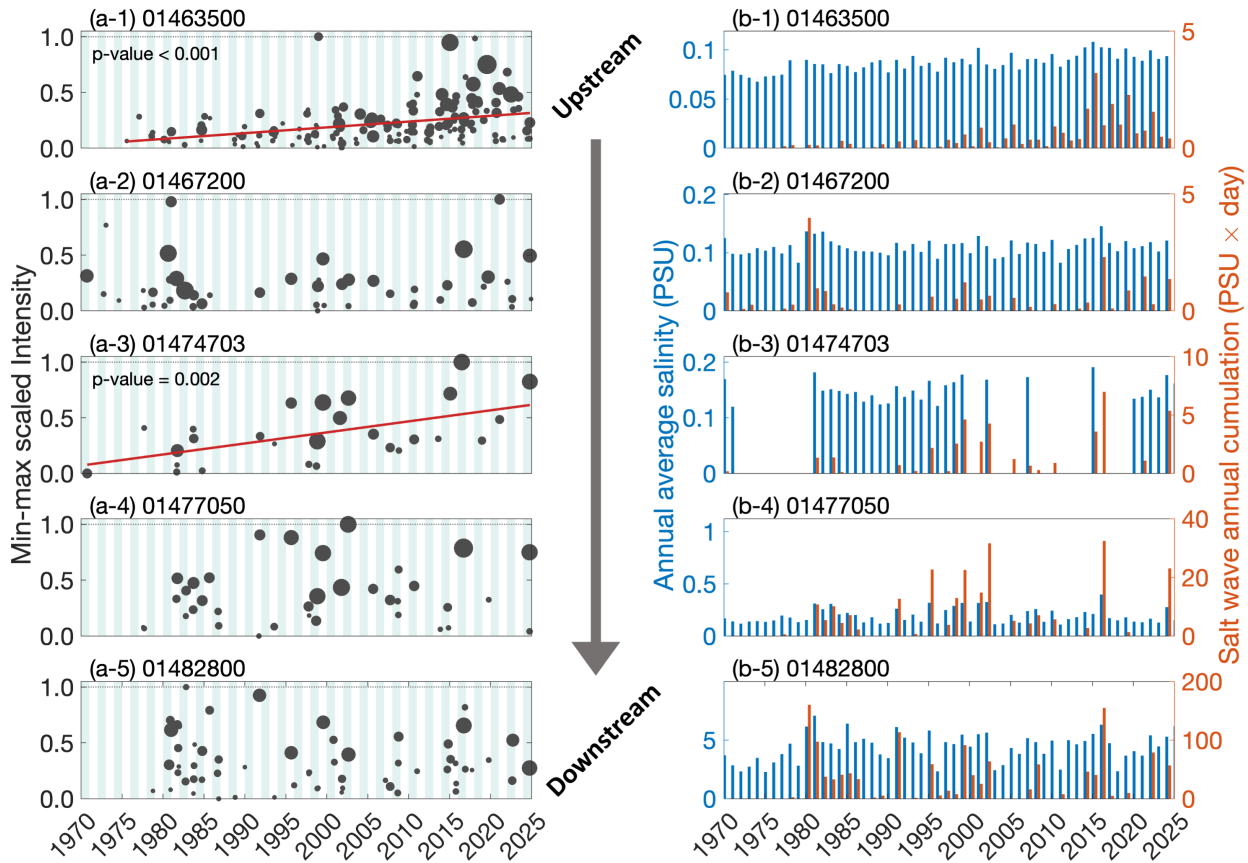

**Figure 2**. (a) Salt wave events detected over multiple decades at five USGS stations along the Delaware Bay, arranged from upstream to downstream. Each circle denotes an individual event, with its size corresponding to the event's duration and its vertical position reflecting the range of salt wave intensities (minimum to maximum) recorded between 1970 and 2024, during periods when all five stations had data. The zebra-striped background emphasizes recurring annual patterns in salt wave occurrences. (b) Annual mean salinity shown as a blue bar plot, with a red bar indicating the total annual cumulation of salt wave events.

3.2 High salinity years do not guarantee severe salt wave events

Along the Mid-Atlantic coast, the magnitude of salinity trends varies dynamically across years. Importantly, a high-salinity year does not necessarily correspond to higher frequency, longer duration, stronger intensity, or larger cumulation of salt wave events. For example, although 2009 was a high salinity year (local 65th percentile) in Delaware Bay, no salt wave events were observed, whereas a comparable year of annual mean salinity, 2015, experienced two distinct salt wave events (Fig. 1ab). Another example is 2017 in the Chesapeake Bay, when annual salinity exceeded the local 70th percentile, only a single salt wave event was recorded at a couple of the lower Bay stations (Supplement Fig. 1). These examples illustrate that high ambient salinity alone is not a reliable predictor of salt wave severity.

### 3.3 Lower salt wave frequency does not always imply lower SWI severity

A lower frequency of salt wave events does not necessarily indicate reduced salinization stress, especially when long-duration events occur. In fact, individual events with prolonged durations often exhibit higher mean intensities, contributing disproportionately to the cumulative salt load (Fig. 3). Across all five salinity zones, such extended events consistently show elevated mean intensities (Supplement Fig. 3). In these cases, the frequency of events becomes a less meaningful indicator of impact, while metrics like mean and cumulative intensity provide a clearer picture of extreme saltwater intrusion (SWI) conditions. These findings highlight that prolonged, high-intensity salt waves – though less frequent – can lead to amplified chronic and osmotic stress on coastal ecosystems, increasing the risk of ecological disruption and long-term shifts in community composition.

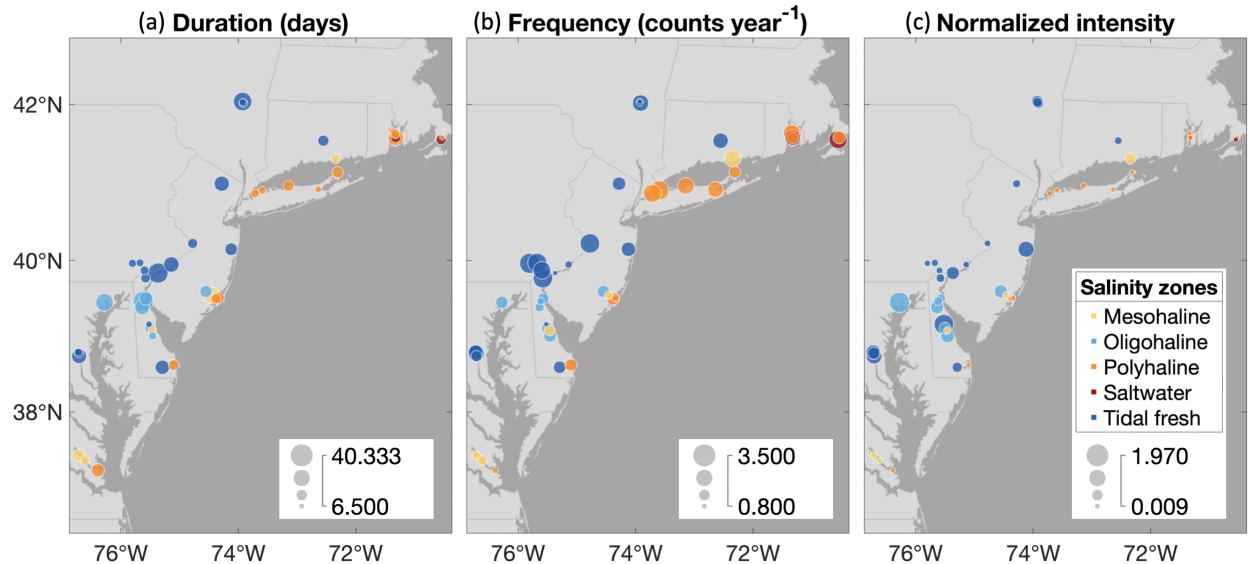

**Figure 3**. Map of the study area covering the mid-Atlantic and parts of the New England coast. Circles represent observation stations, with colors indicating distinct salinity zones based on long-term annual average salinity. Circle size corresponds to the mean salt wave (a) duration, (b) frequency, and (c) normalized intensity relative to the local mean salinity.

3.4 Salinity zone impacts the spatial patterns of salt wave characteristics

Spatially, long-term trends in annual average salinity tend to be consistent among monitoring stations located within the same region or in adjacent salinity zones or hydrologic units (Supplement Fig. 1a). In contrast, adding salt wave metrics, additional interesting patterns emerge (Supplement Fig. 1b). Although salt waves are more likely to co-occur at stations nearby, strong heterogeneity is clearly shown between different salinity zones, even within the same region. Oligohaline and tidal fresh zones typically experience salt waves of longer duration and greater intensity (Fig. 3) while polyhaline and mesohaline zones, due to their closer proximity to seawater, tend to encounter more frequent but milder and shorter salt waves (Fig. 3). These spatial patterns highlight the importance of incorporating salt wave metrics to capture localized dynamics at different salinity zones.

3.5 Localized salt wave events happen within broader systems

While salt wave characteristics are often similar within the same broader system, notable heterogeneity still occurs – particularly in marginal areas. For example, the Murderkill River in lower Delaware Bay (a lower Bay tributary; the 3rd oligohaline station in Supplement Fig. 1) experienced frequent salt wave events in 2012 but fewer in 2017, while the main stem of Delaware Bay (majority of the tidal fresh stations) showed the opposite pattern, with few events in 2012 but many in 2017. In addition to the contrast between the tributaries and the main stem of the Delaware Bay, variability across different salinity zones is also evident along the main stem of the bay – stations display differing trend patterns (Fig. 2). Upstream stations show a clear increase in the frequency and intensity of salt wave events over the past five decades, whereas downstream stations exhibit a much less pronounced trend (Fig. 2). These patterns highlight the spatial complexity of salt wave behavior, showing that even within a single estuarine system, localized salt wave characteristics are shaped by distinct drivers and carry unique ecological implications.

**4 Drivers of coastal salt waves**

Just as long-term SWI trends are shaped by large-scale factors – such as global sea-level rise, land subsidence, groundwater depletion, reduced freshwater discharge, and decadal climate anomalies like the Great Salinity Anomaly (Dickson et al., 1988) – coastal salt waves are also influenced by these persistent forces. In Delaware Bay, over six decades of observations reveal the growing influence of sea-level rise on both chronic SWI and the occurrence of salt wave events. Notably, while average annual salinity has shown only modest increases since the 1970s, the intensity, duration, and frequency of salt waves have risen substantially (Fig. 2). In more recent years, higher levels of SWI or stronger salt waves have been observed, even under similar annual freshwater discharge conditions (Supplement Fig. 2; e.g., year 2001 vs. 2016). These findings indicate that long-term drivers set the stage for SWI and therefore the salt wave events.

However, the occurrence and severity of coastal salt waves are often more closely linked to short-term processes and event-scale variability, including extreme events such as droughts, discharge events, and coastal storms (Table 1). In addition, the dominant drivers of coastal salt waves vary with time and locations. Firstly, drought in a specific watershed can result in saltier estuaries (Soileau et al., 1990), leading to more frequent or intense salt wave events. The spatial patterns observed salt wave events suggest that watershed zonation may partly account for the co-occurrence of salt wave events in the same or neighboring regions (section 3.4), further reflecting the influence of drought conditions. Overall, drought is a key driver of coastal salt waves by amplifying salinity levels in estuaries and contributing to the spatial clustering of events across similarly affected watersheds.

Freshwater discharge often plays a more significant role in influencing salt wave events. For example, freshwater released from the six major reservoirs in the Delaware River Basin helps regulate the salt front in the main stem of the bay and tends to lower the occurrence of salt wave events in the upper Delaware Bay and River (Hammond and Fleming, 2021). However, this management may have played a lesser role in regulating salt wave events in tidal creeks, due to differences in discharge dynamics between the main stem and its tributaries. Overall, these patterns reveal the critical role of flow alteration in shaping salt wave dynamics.

Unlike upstream regions where salt wave events are primarily triggered by variations in freshwater discharge, downstream estuaries are more strongly influenced by coastal storms. These storms can lead to saltwater flooding, potentially causing salt wave occurrences – especially when freshwater runoff from heavy rains is delayed or minimal. Following such storm events, factors like water-level gradients, discharge dynamics, and tidal pumping significantly contribute to the propagation of salt waves (Guerra-Chanis et al., 2021). The response of a given coastal system can vary, but in some cases, storm surge flooding may trigger prolonged salt wave episodes due to the slow recovery of the system from salinization (Keim et al., 2019). Overall, the complex interplay of hydrological and meteorological processes shapes the salt wave dynamics, particularly in storm-affected coastal regions.

Additionally, in soils and groundwater systems, processes such as evapotranspiration, rainfall, flood overtopping, and subsequent flushing play important roles in salinity dynamics (Paldor and Michael, 2021; Cantelon et al., 2022). Unlike surface water systems, these subsurface environments respond to a more complex interplay of hydrologic and climatic factors. For example, during dry conditions, water loss through evapotranspiration can concentrate salts near the root zone (Nordio and Fagherazzi, 2024). In contrast, heavy rainfall can rapidly infiltrate the soil, flushing accumulated salts downward through the profile (Nordio and Fagherazzi, 2024). Additionally, when coastal or riverine floods overtop boundaries, they can introduce saline water that slowly seeps into soils or shallow groundwater, gradually shifting subsurface salinity levels over time (Yu et al., 2024). These distinct mechanisms of salt movement contribute to localized salinity pulses in soil and groundwater systems.

# 5 Conclusions

To better capture salinity anomalies that may significantly disrupt coastal ecosystems, we present the concept of coastal salt waves to capture the variability of extreme SWI events. This framework incorporates multiple dimensions of SWI anomaly events – frequency, duration, timing, intensity, and cumulation – allowing for tailored evaluations of system stress based on the aspect of interest. Cases studies show that the occurrence of salt wave events is independent of mean salinity levels in coastal systems, and an increase in both the intensity and frequency of salt wave events does not necessarily correspond to long-term rises in annual mean salinity. We encourage its application to enhance understanding of the environmental drivers and impacts of climate change on coastal ecosystems. Such insights would be valuable for coastal management and conservation efforts.

**Acknowledgments**

XC is supported by an NSF OCE-PRF fellowship (grant no. 2403359). QQ and JS are supported by the U.S. Environmental Protection Agency (4H95317201). MK and HM are supported by the Coastal Critical Zone Network (National Science Foundation award EAR2012484). PR is supported by the U.S. Department of Energy (DE-SC0024709).

**Open Research**

All the salinity data are publicly available from USGS (https://dashboard.waterdata.usgs.gov/app/nwd/en/) and NERR (https://cdmo.baruch.sc.edu/pwa/index.html). Scripts for case study data analysis can be accessed at a github repository (doi: 10.5281/zenodo.15384511).

**Appendix**

*Section 1. Regional setting and data processing*

The case study area spans across the Long Island Sound, Delaware Bay, Chesapeake Bay, and Delmarva Peninsula – one of the largest coastal plain estuarine systems (Fig. 3). This extensive estuarine waterbody is characterized by large-scale gravitational circulation, resulting in complex salinity dynamics throughout the system (Pritchard, 1952; Geyer et al., 2020). Additionally, this region is recognized as a climate change hotspot, experiencing a relative sea-level rise two to three times faster than the global average (Ezer et al., 2013; Miller et al., 2013). Over recent decades, significant marsh loss and forest retreat have been documented (Kirwan and Gedan, 2019; Mondal et al., 2023). Beyond climate change impacts, human activities and management practices significantly influence this region (Chen and Kirwan, 2022). For instance, six major reservoirs within the Delaware River Basin regulate water diversions to New York City's water supply and release water into the Delaware River for conservation purposes (Hammond and Fleming, 2021). These combined effects of climate change and human activities drive dynamic salinity fluctuations, including salt wave events, making this region an ideal case study for evaluating salt wave metrics for future monitoring and management.

We analyzed salinity and specific conductivity data at surfaces water stations using datasets from the US Geological Survey (USGS; https://dashboard.waterdata.usgs.gov/app/nwd/en/) and the National Estuarine Research Reserve System (NERRS; https://cdmo.baruch.sc.edu/pwa/index.html). Most USGS stations provide data only after 2007 and most NERRS station data are available after 2004. However, along the main stem of Delaware Bay, observational records can extend as far back as the 1960s. For stations reporting specific conductance, we applied the MATLAB toolbox (version 3_06_16) based on the Thermodynamic Equation of Seawater – 2010 (TEOS-10; https://github.com/TEOS-10/GSW-C) to calculate salinity in PSU if needed. All time series were filtered into daily averages to identify salt wave events and estimate their duration and intensity. Only stations with at least 10 years of observation were included in the salt wave analysis. Additionally, to account for incomplete annual measurements, we required a minimum of 70% data availability within a year to include it in the analysis of salt wave events. To provide an overview of the annual salinity trend across stations in different salinity zones, we normalized the annually averaged salinity using the min-max scale over the local records (Jain and Bhandare, 2013).

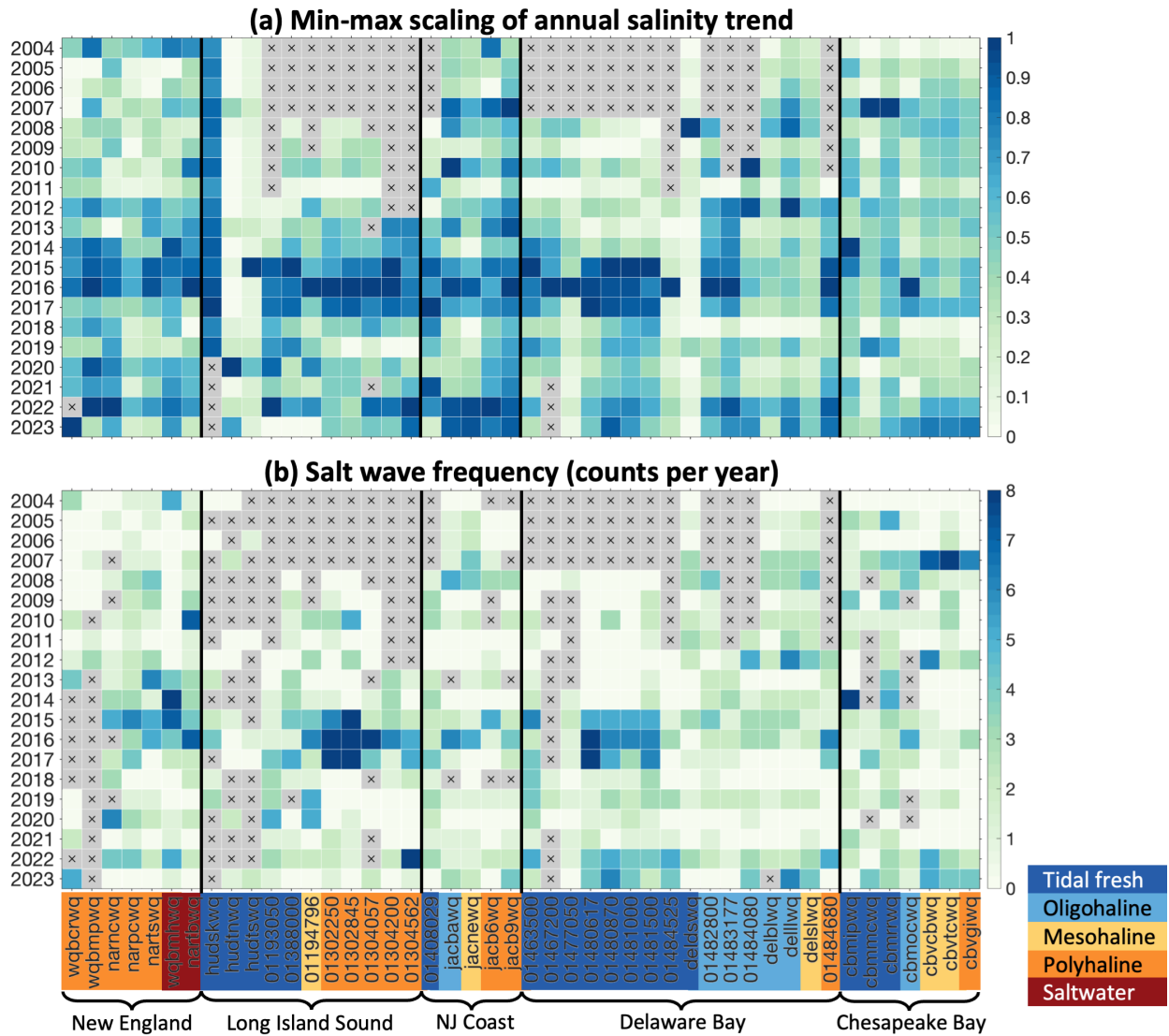

**Supplement Figure 1**. (a) Min-max scaling of salinity trends and (b) annual salt wave frequency counts for each station along the mid-Atlantic and partial New England coasts as shown in the map at Supplement Fig. 3e. Grey squares with crosses indicate stations with insufficient observational data. In panel (a), colors transitioning from white to blue represent increasing local salinity trends. In panel (b and c), the color of each station name reflects its classification based on the mean local salinity record.

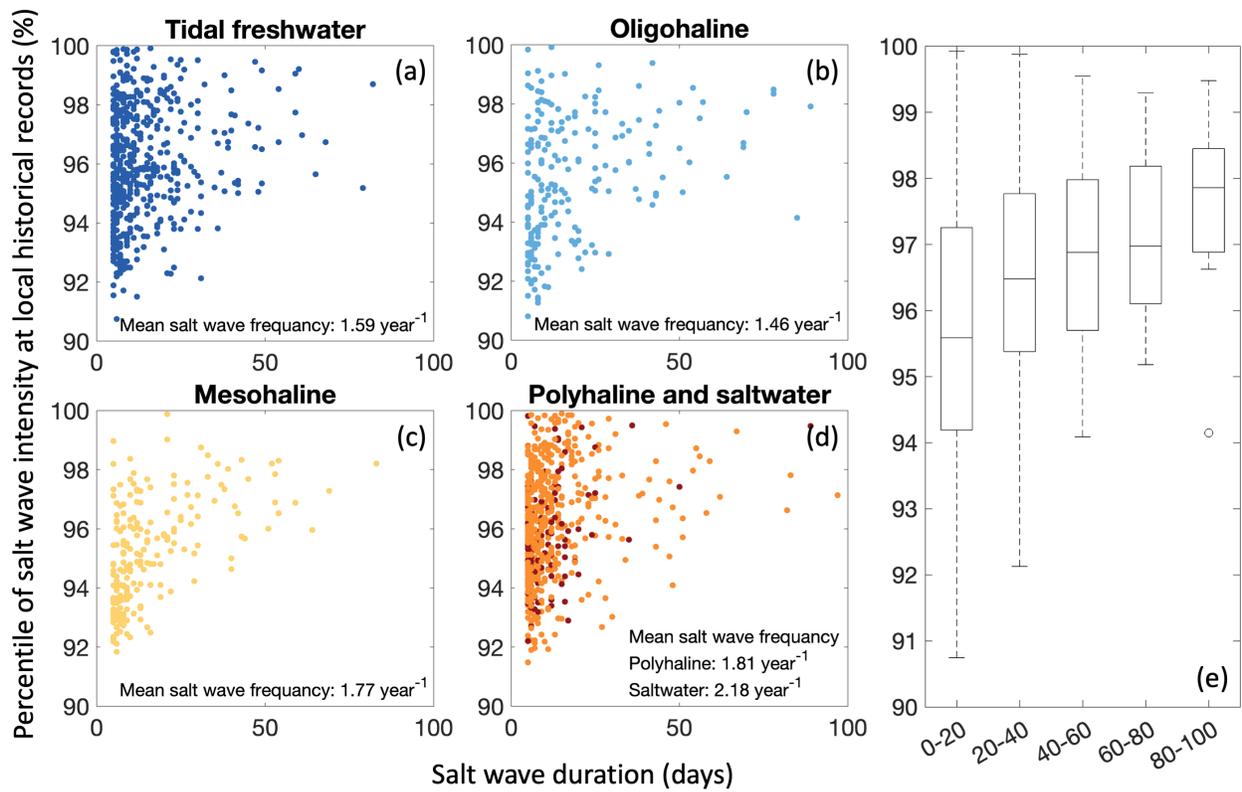

**Supplement Figure 2**. (a-d) Salt wave duration plotted against the normalized intensity of each event, with (e) box plot across all the salinity zones. Intensities have been standardized relative to percentiles in the local historical records.

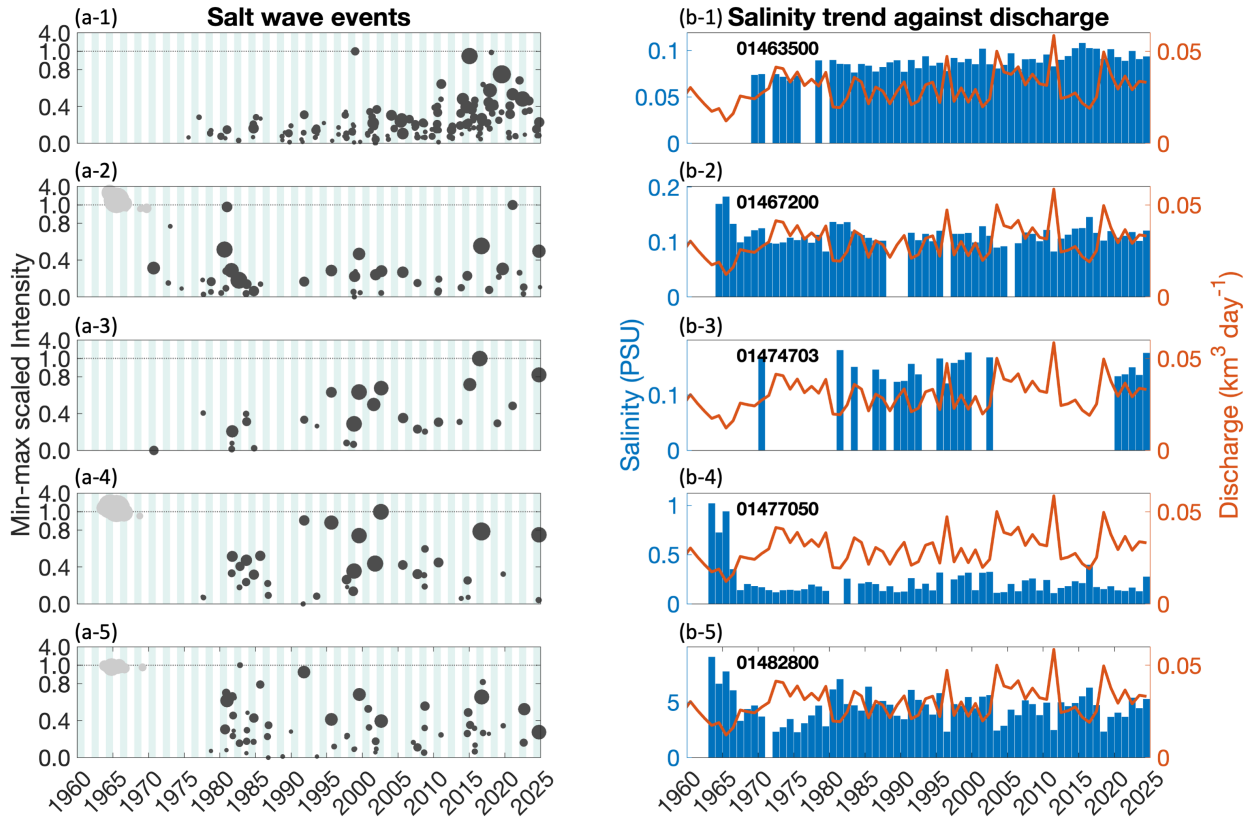

**Supplement Figure 3**. (a) Salt wave events identified in a six-decade dataset from 5 USGS station along the Delaware Bay. Each circle represents a single event, with its size proportional to the event duration and its position on the y-axis indicating the min-max scale of the salt wave intensities over 1970 to 2024 when all 5 stations has observations. The zebra-striped background highlights annual patterns accounting of salt wave events. The light gray circles represent the magnitude of salt wave events at certain stations while no records are available all these stations during in the 1960s. (b) Annual mean salinity represented as a bar plot, with the red line indicating annual discharge from the available historical record at USGS station 01463500, upstream of the Delaware Bay.

*Reference cited in Appendix*